\documentclass[a4paper,fleqn,useAMS,usenatbib]{mnras}

\pdfoutput=1

\usepackage{mathptmx}
\usepackage[T1]{fontenc}
\usepackage{ae,aecompl}

\usepackage[dvips]{graphicx}
\usepackage{amsmath}
\usepackage{amsfonts}
\usepackage{amssymb}
\usepackage{comment}
\usepackage{pdflscape}
\usepackage[dvipsnames]{xcolor}
\usepackage[%
        ]{hyperref}
        
\hypersetup{
	colorlinks=true,	% false: boxed links; true: colored links
	urlcolor=MidnightBlue,		% color of external links
	pdfpagelabels=true,
	hypertexnames=true,
	plainpages=false,
	naturalnames=true,
	pdftitle={Forecasted planetary radii},    % title
	pdfauthor={Yi et al.},     % author
	linkcolor=WildStrawberry,          % color of internal links (change box color with linkbordercolor)
	citecolor=ForestGreen,        % color of links to bibliography
}

\newcommand{\kepler}{{\it{Kepler}}}

\newcommand{\forecaster}{{\tt forecaster}}

\newcommand{\actuallink}{https://github.com/CoolWorlds/cheopsforecasts}
\newcommand{\httplink}{\href{\actuallink}{this URL}}

\title[Forecasted planetary radii]{Forecasting the detectability of known radial velocity
planets with the upcoming CHEOPS mission}
\author[Yi et al.]{Joo Sung Yi$^{1}$\thanks{E-mail:
\href{mailto:dyi@taftschool.org}{dyi@taftschool.org}}, Jingjing Chen$^{2}$ \& David Kipping$^{2}$\\
$^{1}$Taft School, 110 Woodbury Rd, Watertown, CT 06795 \\
$^{2}$Dept. of Astronomy, Columbia University, 550 W 120th Street, New York NY 10027
}

\date{Accepted . Received ; in original form }

\pubyear{2017}

\begin{document}
\label{firstpage}
\pagerange{\pageref{firstpage}--\pageref{lastpage}}
\maketitle

\begin{abstract}
The \textit{Characterizing Exoplanets Satellite} (CHEOPS) mission is planned
for launch next year with a major objective being to search for transits of
known RV planets, particularly those orbiting bright stars. Since the radial
velocity method is only sensitive to planetary mass, the radii, transit
depths and transit signal-to-noise values of each RV planet are, a-priori,
unknown. Using an empirically calibrated probabilistic mass-radius relation,
\forecaster\ \citep{chen:2017a}, we address this by predicting a catalog
of homogeneous credible intervals for these three keys terms for 468 planets
discovered via radial velocities. Of these, we find that the vast majority should be
detectable with CHEOPS, including terrestrial bodies, if they have the correct
geometric alignment. In particular, we predict that 22 mini-Neptunes and 82
Neptune-sized planets would be suitable for detection and that more than 80\%
of these will have apparent magnitude of $V<10$, making them highly suitable
for follow-up characterization work. Our work aims to assist the CHEOPS team
in scheduling efforts and highlights the great value of quantifiable,
statistically robust estimates for upcoming exoplanetary missions.
\end{abstract}


\section{Introduction}
\label{sec:intro}

The \textit{Kepler Mission} has been an extraordinary success for discovering
new exoplanets \citep{coughlin:2016} and constraining the occurrence rate of
planetary companions to other stars \citep{burke:2015,dressing:2015}.
Specifically, \kepler\ identified previously unsuspected features of planetary
systems, including the abundance of planets between the size of Earth and
Neptune \citep{howard:2012}, and a population of very compact multiple-planet
systems \citep{lissauer:2011}, as well as the discovery of transiting
circumbinary planets \citep{doyle:2011}.

Despite these successes, a downside to \kepler\ is that the majority of its
worlds orbit relatively faint host stars, with the median host star \kepler\
apparent magnitude of a planetary candidate being 14.6 at the time of writing
(see \href{http://exoplanetarchive.ipac.caltech.edu}{NASA Exoplanet Archive},
\citealt{akeson:2013}). This poses a severe challenge for a variety of
follow-up observations, including collecting sub m\,s$^{-1}$ radial velocities
(RVs) and atmospheric characterization. In order to provide transiting planets
orbiting brighter stars, which would be more suitable for follow-up work,
several on-going and near-future transit surveys aim to survey a larger
fraction of the sky, such as HATNet \& HATSouth \citep{bakos:2004,bakos:2013},
WASP \& WASP-South \citep{pollacco:2006,wilson:2008}, MEarth \& MEarth South
\citep{charbonneau:2009,irwin:2015}, TESS \citep{ricker:2015}, NGTS
\citep{wheatley:2013} and PLATO \citep{rauer:2014}, to name a few.

An orthogonal approach to discovering bright, transiting planet is not to
conduct a wide-field, blind transit survey but to target specific stars known
to already harbor RV planets. Such an approach has been successful in the
past, in particular with the MOST telescope \citep{rucinski:2003}, which
discovered transits of 55 Cancri e \citep{winn:2011} and HD 97658b
\citep{dragomir:2013}, as well as the first transit search for Proxima b
\citep{kipping:2017}. Although MOST was certainly not dedicated to transit
follow-up of RV planets, these successes have highlighted the high impact that
a dedicated, small space-based telescope could have on the field. It is
therefore of no great surprise that such a mission is now expected to fly in
2018 \citep{broeg:2013}, the \textit{Characterizing Exoplanets Satellite}
(CHEOPS).

For each target pursued, the CHEOPS team will, in general, only know the
minimum mass of any planets in orbit, their ephemeris and properties of the
host star. Notably the actual radius and corresponding transit depth expected
are, a-priori, unknown. Since the transit depth directly controls the
signal-to-noise ratio (SNR), there is clearly a need to estimate such a term
reliably. Pursuing targets for which the SNR will be too small to
confirm a detection is naturally an ineffective use of the limited observing
time.

The minimum mass provides the best proxy for estimating planetary radius.
Indeed, it is not unreasonable to simply treat the assumed mass to be equal to
the minimum mass, since this will, at worst, lead to a conservative,
under-estimate of the true SNR for all planets except
self-compressed brown dwarfs (this is discussed in more detail in later in
Section~\ref{sub:ministrue}). Nevertheless, converting from
mass to radius is not a straight-forward enterprise, since the relation is
intrinsically probabilistic, a result which was both expected theoretically
\citep{seager:2007,fortney:2011,batygin:2013} and has been established
empirically \citep{wolfgang:2016,chen:2017a}. Nature builds a diverse mixture
of planets at any fixed size or mass, reflecting the differing compositions,
histories and environments from which they originate. This means that naively
invoking deterministic mass-radius relations when forecasting radii will
underestimate the a-posteriori credible interval, but can also be
systematically biased for planets which reside near critical transitional
boundaries, such as the Terran-Neptunian boundary (this is discussed further
in Section~\ref{sub:results}), where the mixture model
nature of the posterior is decidedly non-Gaussian \citep{chen:2017b}.

To address this, this short paper presents forecasted credible intervals for
the radius, transit depth and corresponding CHEOPS signal-to-noise for all
known RV planets using the empirical, probabilistic \forecaster\ model of 
\citet{chen:2017a}. In Section~\ref{sec:forecasting}, we describe how we
predict radii from masses, including a careful propagation of measurement
uncertainties, for each RV planet. In Section~\ref{sec:results}, we
highlight some important results and implications from this effort before
discussing future work and limitations of our results in
Section~\ref{sec:discussion}.

\section{Forecasting Radii from Masses}
\label{sec:forecasting}

\subsection{Probabilistic predictions with \forecaster}
\label{sub:forecaster}

A basic requirement of this work is to predict the transit depth, and thus
radius, of each of the few hundred exoplanets discovered through the radial 
velocity method. A major challenge facing any attempt to convert masses to
radii, or vice versa, is that the mass-radius relation of exoplanets displays
intrinsic spread, as outlined in Section~\ref{sec:intro}. This spread,
likely due to intrinsic variations in composition, chemistry, environment,
age and formation mechanism, means that simple deterministic models are
unable to reliably predict the range of plausible radii expected for any
given mass.

One solution to this problem is to model the exoplanetary masses and radii
with a probabilistic relation. This treats each mass as corresponding to
a probability distribution of radii, rather than a single, deterministic
estimate. \citet{chen:2017a} inferred such a relation for masses spanning
nine orders-of-magnitude, providing both a comprehensive scale for inversion,
as well as calibrating the relation as precisely as possible by utilizing
the full dynamic range of data available. An additional key quality of
the \forecaster\ model, as \citet{chen:2017a} refer to their model as,
is that the relationship is trained on the actual measurement posteriors of
each training example via a hierarchical Bayesian model, meaning that the
measurement error of these data are propagated into the \forecaster\
predictions.

For these reasons, \forecaster\ is a natural tool for predicting the radii
of radial velocity planets in what follows\footnote{Indeed this was actually
highlighted as one of the intended uses of the code in the original paper.}.
We highlight that \forecaster\ has already been utilized in the reverse
application, predicting an ensemble of exoplanet masses from radii in the
follow-up paper of \citet{chen:2017b} for \kepler\ planetary candidates.

We direct the reader to \citet{chen:2017a} for details in the mathematical
model, hierarchical Bayesian inference and regression tools used to build
\forecaster. In what follows, we focus instead on our implementation.

\subsection{Accounting for measurement uncertainties}

Whilst \forecaster\ accounts for the intrinsic dispersion displayed by nature
in the mass-radius relation, as well as the measurement uncertainties of
its training set, a third source of error exists that requires accounting for
on our end in this work. Specifically, for each RV planet, there exists often
sizable uncertainty in the mass of each body.

To account for this, we technically require the a-posteriori probability
distribution of each planet's mass, $m$. For $n$ samples randomly drawn from
such a mass posterior, $\mathrm{Pr}(m)$, a robust forecast of the planetary
radius can be made by proceeding through the samples, row-by-row, and
executing \forecaster\ at each line. The compiled list of predicted radii
represents the covariant posterior distribution for the forecasted radius. A
similar process is utilized in the reverse-direction application described in
\citet{chen:2017b}.

A practical challenge to implementing the scheme above is that posterior
distributions are rarely made available in papers announcing or studying
RV planets\footnote{This is in contrast to our earlier work predicting masses
from radii in \citet{chen:2017b}, where transit-derived posteriors are often
available.}. Fortunately, of all the parameters describing the RV model, the RV
semi-amplitude ($K$) posterior rarely exhibits multi-modality or extreme
covariance \citep{ford:2006} and can often, in practice, be reasonably
approximated as Gaussian (e.g. see \citealt{tuomi:2012} and
\citealt{hou:2014}). By extension, since $K \propto m$ for $m \ll 
M_{\star}$, then we will assume that the planetary mass can be described as
Gaussian, such that $m \sim \mathcal{N}[\mu_m,\sigma_m]$, in what follows in
order to make progress.

\subsection{Approximate form for the posterior distributions}

Naively using Gaussians for $m$ can be problematic though for two reasons.
First, Gaussians have non-zero probability density at negative values and
thus negative masses will occasionally be sampled from a Gaussian distribution.
Second, Gaussians are perfectly symmetric yet literature quoted credible
intervals for $m$ may include asymmetric uncertainties e.g. $m = 
(\mu_m)_{ -\sigma_{m-} }^{ +\sigma_{m+} }$. In practice, we note that none
of the reported planetary masses used for this work were asymmetric,
although several credible intervals associated with stellar properties were
and thus the problem requires solving regardless.

The first issue can be tackled by invoking a truncation of a Gaussian
distribution, preventing the distribution from drawing negative samples. This
is most readily achieved by truncating from $m=0$ up to $m=\infty$.
Accordingly, the probability density function (PDF) is modified from

\begin{align}
\mathrm{Pr}(m;\mu_m,\sigma_m) &= \frac{1}{\sqrt{2\pi}\sigma_m} \mathrm{exp}\Bigg( \frac{(m-\mu_m)^2}{2\sigma_m^2} \Bigg)
\end{align}

to

\begin{equation}
\mathrm{Pr}(m;\mu_m,\sigma_m) =
\begin{cases}
\frac{1}{1-\tfrac{1}{2}\mathrm{erf}[\tfrac{\mu_m}{\sqrt{2}\sigma_m}]} \frac{1}{\sqrt{2\pi}\sigma_m} \mathrm{exp}\Big( \frac{(m-\mu_m)^2}{2\sigma_m^2} \Big)  & \text{if } m \geq 0,\\
0 & \text{if } m < 0 .
\end{cases}
\end{equation}

The second issue can be solved in a number of ways but here we tackle it by
modifying our posterior to be a mixture model of two truncated Gaussians.
Specifically, the region from $0 \leq m < \mu_m$ is handled by one component
modeling the negative uncertainty direction, and the region $\mu_m \leq m <
\infty$ is handled by a second modeling the positive uncertainty direction.
The negative component is a Gaussian distribution given by
$\mathcal{N}[\mu_m,\sigma_{m-}]$ and truncated to the region $[0,\mu_m]$,
whereas the positive component is $\mathcal{N}[\mu_m,\sigma_{m+}]$ and
truncated to the region $[\mu_m,\infty]$. Since the Gaussians have different
variances, they will, in general have distinct densities at the meeting point
of $m=\mu_m$ and so we force the densities to be equal by setting the mixture
weights accordingly. The final approximate form for the mass posterior
distribution is thus taken to be

\begin{equation}
\mathrm{Pr}(m;\mu_m,\sigma_{m-},\sigma_{m+}) =
\begin{cases}
\frac{1}{\zeta_{-}} \mathrm{exp}\Big( \frac{(m-\mu_m)^2}{2\sigma_{m-}^2} \Big)  & \text{if } 0 \leq m < \mu_m,\\
\frac{1}{\zeta_{+}} \mathrm{exp}\Big( \frac{(m-\mu_m)^2}{2\sigma_{m+}^2} \Big)  & \text{if } \mu_m \leq m < \infty,\\
0 & \text{if } m < 0,
\end{cases}
\end{equation}

where

\begin{align}
\zeta_{-} &= \pi \sigma_{m-} \sigma_{m+}
\Bigg( \frac{1}{\sqrt{2\pi}\sigma_{m-}} + \frac{1}{\sqrt{2\pi}\sigma_{m+}} \Bigg)
\mathrm{erf}\Big[ \frac{\mu_m}{\sqrt{2}\sigma_{m-}} \Big]
\end{align}

and

\begin{align}
\zeta_{+} &= \pi \sigma_{m-} \sigma_{m+}
\Bigg( \frac{1}{\sqrt{2\pi}\sigma_{m-}} + \frac{1}{\sqrt{2\pi}\sigma_{m+}} \Bigg).
\end{align}

In practice, for each planet, we draw $n=10^5$ posterior samples for the
planetary mass, which are then passed through to \forecaster. It is easy to
verify the above probability density integrates to unity when marginalized over
$m$.

\subsection{Treating minimum mass as being equal to true mass}
\label{sub:ministrue}

As an aside, we highlight that in what has been described thus far, and indeed
what follows throughout, we assume that the minimum mass equals the true mass,
in other words $\sin i \simeq 1$. Whilst one might naively assume $\sin i$
should be, a-priori, isotropically distributed, our work primarily
concerns itself with deriving the posterior distribution of the transit SNR
conditioned upon the assumption that the planet transits i.e.
$\mathrm{Pr}(\mathrm{SNR}|\hat{b})$, where $\hat{b}$ denotes that the planet
transits to follow the notation of \citet{sandford:2016}. Within the range of
inclination angles which lead to transits, we can safely assume $\sin i \simeq
1$ to be an excellent approximation. The a-priori probability that the
condition of a transit is satisfied is treated separately by computing transit
probabilities, as described in Section~\ref{sub:tranprob}.

We also highlight that the mass-radius relation is
well-described by a power-law which displays close to monotonic behavior.
The only negative index occurs in the degenerate regime of Jovian worlds, due
to gravitational self-compression \citep{chen:2017a}, although the effect is
comparatively small. Accordingly, by assuming that the true mass equals the
minimum mass, in general we end up underestimating the forecasted radius, which
in turn means that we underestimate the expected CHEOPS transit SNR. If the
SNR is underestimated, then CHEOPS will still be expected to achieve a
detection, whereas overestimates are far more problematic leading to
potentially wasted telescope time. As noted in the previous paragraph
though, in practice the minimum SNR derived will be extremely close to
the true value since $\sin i \simeq 1$ in order to satisfy the conditional that
the planet transits.

In summary, by assuming the true mass equals the minimum mass, we introduce
a very small underestimate of the true SNR, but this is certainly
acceptable for the purposes of assigning telescope resources, for the reasons
described above.

\subsection{A homogeneous data source}
\label{sub:data}

Having established the procedure by which radius forecasts will be computed,
we require a curated list of literature-derived credible intervals for each
planetary mass. Rather than curate a list ourselves, we turn to the 
``Exoplanet Orbit Database'' (\href{http://exoplanets.org}{EOD};
\citealt{han:2014}) resource. EOD represents ``a carefully constructed
compilation of quality, spectroscopic orbital parameters of exoplanets orbiting
normal stars from the peer-reviewed literature'' and is an ideal resource
for the purposes of this paper.

Literature reported planetary mass credible intervals were obtained for a total
of 481 RV planets from the EOD using the filter {\tt PLANETDISCMETH = RV} on the
$29^{\mathrm{th}}$ August 2017. Amongst this sample, however, we encountered a
few cases where the standard errors on $m$ were missing, and these objects were
ignored in what follows (eta Ceta b \& c, 91 Aqr A b, and HD 187085b), leaving
us with a total of 477 RV planets.

\subsection{Stellar radii posteriors}
\label{sub:stars}

An additional advantage of using the EOD is that for each planet, there is
also, in general, an associated set of stellar parameters. Although not
necessary for predicting the radii of the 477 RV planets, stellar radii
posteriors of the host stars are important for calculating the predicted
transit depths, and in turn the CHEOPS SNRs.

During our work, we noted that 33 planets were missing both a
maximum a-posteriori estimate for their parent star's radius, as well as an
associated uncertainty. However, we noted that all of the objects has reported
stellar masses in the EOD. In order to assign a stellar radius in these
instances, we elected to again use \forecaster\ to predict the corresponding
radii.

However, for 9 of these 33 planets missing stellar radii, even \forecaster\
was untenable, since their maximum a-posteriori reported stellar masses
exceeded the calibration range upon which \forecaster\ was trained.
Specifically, any mass exceeding $0.87$\,$M_{\odot}$ cannot be asserted to
have a main sequence lifetime less than the present age of the Universe, and
thus stars beyond this mass may have ascended the giant branch already. For
such stars, the intrinsic spread in the mass-radius relation becomes enormous
and \citet{chen:2017a} simply ignore this regime and forbid \forecaster\ to
make predictions beyond this point. Accordingly, for these 9 stars
(HD 33564, 1.25\,$M_{\odot}$;
HD 48265, 0.90\,$M_{\odot}$; 
HD 132406, 1.09\,$M_{\odot}$;
HD 143361, 0.90\,$M_{\odot}$;
HD 175167, 1.10\,$M_{\odot}$;
HD 204313, 1.02\,$M_{\odot}$;
HD 212301, 1.05\,$M_{\odot}$;
HD 240237, 1.69\,$M_{\odot}$;
HD 132563 B, 1.01\,$M_{\odot}$), we were unable to make any kind of prediction
for their stellar radius and thus dropped them from our list, reducing
the sample from 477 to 468 RV planets.

Of the remaining 24 (33 minus 9) RV planets without stellar radii but
reported masses below $0.87$\,$M_{\odot}$, all of the stars has symmetric
measurement uncertainties on their stellar masses with the exception of
GJ 433 b, GJ 674 b, GJ 667 C b \& c, HD 164604 b and HIP 79431 b (6 planets),
which lacked any reported stellar mass errors at all\footnote{The 24-6-18
aforementioned planets are flagged with a ``1'' later in Table~\ref{tab:final},
whereas the exceptional 6 planets are flagged with a ``2''.}. To assign an
error, we computed the median percentage error on stellar mass for the rest of
the sample, found to be 4.64\%, and thus elected to simply assign a 5\% flat
rate percentage on stellar mass for these 6 objects. Using \forecaster, we
generated the stellar radii posteriors using $10^5$ samples.

Finally, we found that 11 RV planets in our sample had a reported maximum
a-posteriori stellar radius but no associated uncertainty. For these 11
cases\footnote{These 11 cases are flagged with a ``3'' later in 
Table~\ref{tab:final}.}, we adopted a symmetric uncertainty for all equal to
the median percentage error on stellar radii from the sample of planets which
had EOD-reported stellar radii and uncertainties, which was 3.38\% (which we
rounded to 3.5\%).

With a full list of planetary radii and stellar radii posteriors in hand,
the two distributions are multiplied and self-producted to give the
geometric transit depth posterior, $(R_P/R_{\star})^2$.

\subsection{Signal-to-noise ratios}
\label{sub:SNR}

\begin{figure}
\begin{center}
\includegraphics[width=8.4cm,angle=0]{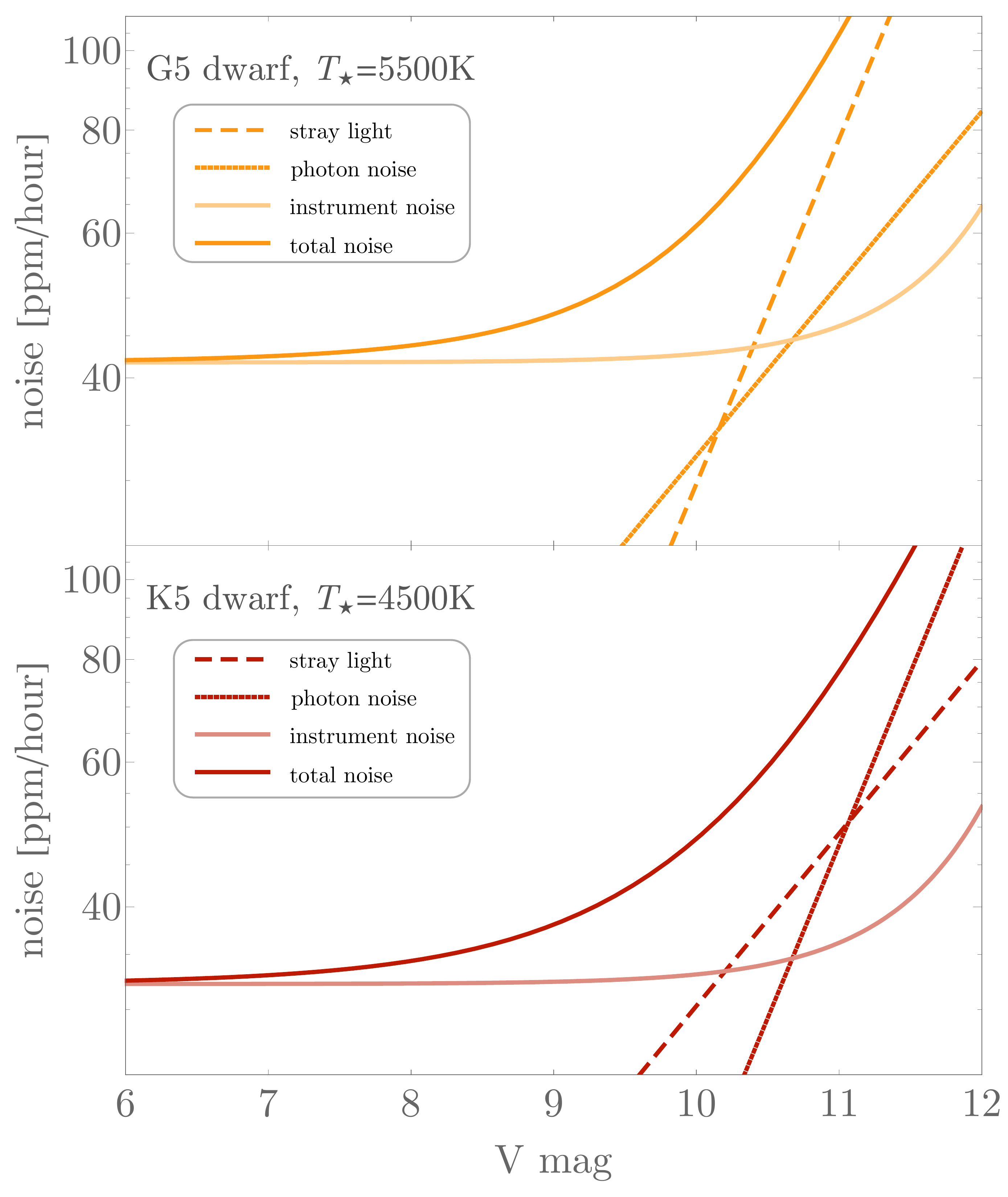}
\caption{
Illustration of the noise model adopted in this work for CHEOPS, showing the
different noise components affecting the total. We linearly interpolate the
G5 and K5 total noise models as a function of effective temperature, except
for temperatures exceeding these two extremes, beyond which we simply adopt the
nearest neighbor model.
}
\label{fig:noise}
\end{center}
\end{figure}

The penultimate step in our calculation is to take the transit depths and
estimate the signal-to-noise ratios (SNRs), as would be seen by CHEOPS.
Courtesy of the CHEOPS team during the preparation of this paper (D.
Ehrenreich; priv. comm.), we were provided with two figures plotting the
expected CHEOPS noise levels from stray light, photon noise and instrument
noise as a function of $V$-band magnitude. The quadrature sum of these three
components defines the total expected noise, providing a more up-to-date noise
model than that described in \citet{broeg:2013}. This is illustrated in
Figure~\ref{fig:noise} where we have reproduced this noise model for a
one-hour cumulative integration time (scaled down from the original
versions we were sent which were for 3\,hours and 6\,hours).

Because of the specific bandpass of the mission, stars of different temperature
will have distinct noise curves and thus we were provided with noise estimates
for a 4500\,K K5-dwarf and a 5500\,K G5-dwarf, as a point of comparison.
Although we do not have a bandpass available, we can estimate the noise at
intermediate temperatures between these two extremes by simply linearly
interpolating between the two, as a first-order approximation. For stars cooler
than 4500\,K, we simply adopt the K5 values and similarly for stars hotter than
5500\,K the G5 model is assumed.

With the noise in hand, one may now easily calculate the SNR over one hour of
cumulative integration time, a term we write as $\mathrm{SNR}_{\mathrm{hour}}$.
For each target, we make this calculation on a posterior sample-by-sample basis
allowing us to predict a posterior for $\mathrm{SNR}_{\mathrm{hour}}$ and
thus the credible intervals, which are reported later in Table~\ref{tab:final}.

It is important to stress that that our one-hour cumulative SNR does not make
any assumptions about how the one hour of integration time is achieved (e.g.
combining multiple orbits or epochs). In practice, the orbit of CHEOPS will
lead to observing gaps and thus one hour of wall time will not equal one
hour of cumulative integration time, unless the target happens to lie within
a continuous viewing zone. We therefore stress that observers will need to
account for the observational duty cycle, $\mathcal{D}$ (based on orbital
viewing and scheduling constraints), the number of transit epochs observed,
$N$, and transit duration, $\tilde{T}$, and the amount of out-of-transit
data acquired, $B$. Accordingly, the actual SNR will follow

\begin{align}
\mathrm{SNR} &= \sqrt{ \sum_{i=1}^N \mathrm{SNR}_{\mathrm{hour}}^2 \Big(
\frac{ \mathcal{D}_i \tilde{T} B_i }{ B_i + \tilde{T} } \Big) },
\label{eqn:SNRreal}
\end{align}

where $\tilde{T}$ is the 1.5-to-3.5 contact transit duration defined in
\citet{investigations:2010}, $\mathcal{D}_i$ and $B_i$ are the duty cycle and
out-of-transit baseline of the $i^{\mathrm{th}}$ transit epoch observed,
amongst $N$ transits. Equation~\ref{eqn:SNRreal} comes from Equation~9 of
\citet{sandford:2016}, and assuming $W \delta \ll T_{14}$, where $W \equiv
(T_{14}+T_{23})/2$ and $T_{14}$ and $T_{23}$ are the first-to-fourth and
second-to-third contact duration. Equation~\ref{eqn:SNRreal} also assumes
that $T_{14} \simeq W \simeq \tilde{T}$ for the sake of simplicity.

Of course, without a detailed orbital model or list of scheduling requirements
imposed by other programs, it is not generally possible at this stage for
us to estimate all of the above parameters in advance. However, we do calculate
a one-sigma credible interval for $\tilde{T}$ in the optimistic limit of an
equatorial transit, accounting for the known eccentricity of the planet using
Equation~15 of \citet{investigations:2010}, which is provided later in
Table~\ref{tab:final}. Nevertheless, our provision of
$\mathrm{SNR}_{\mathrm{hour}}$ offers a straight-forward means for the
community to estimate and rank planets by their detectability under various
circumstances.

\subsection{Transit probabilities}
\label{sub:tranprob}

In what has been described so far, we have computed the probability
distribution of the $\mathrm{SNR}_{\mathrm{hour}}$ of each RV planet as
observed by one integrated hour of CHEOPS telescope time, conditioned upon the
assumption that a planet transits. This may be broadly described as computing
the detectability, the primary goal of this work, but not the actual yield.
The actual yield will be affected by which planets are observed and for how
long, the observing strategy and cadence, and the a-priori transit probability
of each planet. Since the observing strategy for CHEOPS has not yet been
decided upon and will likely evolve during the mission, it is generally not
possible to model the yield at this time. Instead, for context, we compute here
the a-priori transit probability of each planet.

Finally, for each predicted transit depth, we estimate the geometric transit
probability. This calculation is done in the simple case of assuming an
isotropic inclination prior, although we note that the true distribution
is likely more favorable than this but requires detailed modeling of the
mass functions \citep{stevens:2013}. Also, we do not estimate a credible
interval for the transit probabilities, rather simply adopt the maximum
a-posteriori system parameters to compute a single point-estimate for each
planet. This is largely motivated by the lack of publicly available
joint posteriors on eccentricity and argument of periastron, necessary
for such a calculation. Since our $\mathrm{SNR}_{\mathrm{hour}}$ posteriors are
conditioned upon the assumption of a transit, these unknown covariances do not
propagate into our earlier $\mathrm{SNR}_{\mathrm{hour}}$ calculations
fortunately.

In calculating the transit probability, we account for the effects of
orbital eccentricity, $e$ and argument of periapsis, $\omega$, by
using \citep{barnes:2007,burke:2008}:

\begin{equation}
\mathrm{Pr}(\mathrm{transit}) = \Bigg( \frac{R_{\star}}{a} \Bigg) \Bigg( \frac{1+e\sin\omega}{1-e^2} \Bigg).
\end{equation}

In some cases, the eccentricity or argument of periastron was not available
on EOD, in which case simply assume a circular orbit for simplicity. In cases
where $(a/R_{\star})$ was not available, we quote the transit probability as
``N/A'', rather than completely dropping them from our list.

\section{Detectable Planets with CHEOPS}
\label{sec:results}

\subsection{Overview of results}
\label{sub:results}

\begin{figure*}
\begin{center}
\includegraphics[width=17.0cm,angle=0]{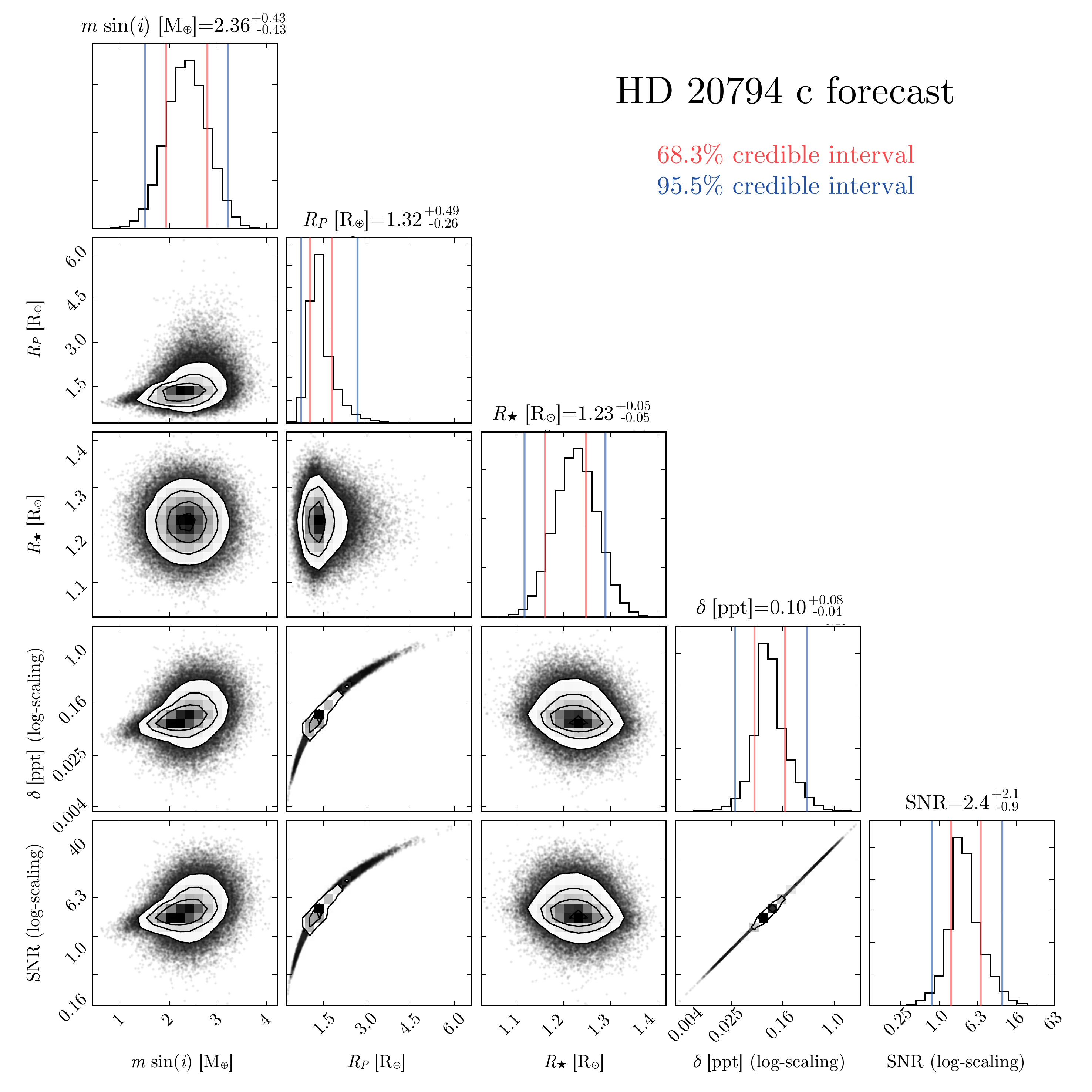}
\caption{
Corner plot of our resulting posteriors for HD 20794c an illustrative
example. Covariances are evident between the parameters, including between
mass and radius.
}
\label{fig:example}
\end{center}
\end{figure*}

After predicting posterior distributions for $R_P$, $\delta$ and
$\mathrm{SNR}_{\mathrm{hour}}$ as described earlier, we tabulated the medians,
as well as the 68.3\% and 95.5\% credible intervals in Table~\ref{tab:final}.
To illustrate a specific example, we show a corner plot of the posteriors
produced in this work for the planet HD 20794c in Figure~\ref{fig:example},
which highlights the covariances, even between mass and radius owing to HD
20794c's location near the Terran-Neptunian divide.

\begin{table*}
\label{tab:final}
\caption{
Final predicted properties for 468 RV planets using \forecaster. Only a portion of the table is shown here, the full version is available in the online version and at \httplink. SNR values correspond to one cumulative hour of integration time.
}
%\centering % used for centering table
\begin{tabular}{lllllll} % centered columns (7 columns)
\hline
Name & $m$\,$[M_{\oplus}]$ & $R_P$\,$[R_{\oplus}]$ & $\delta=(R_P/R_\star)^2$ [ppt] &
%$V$-mag & $\sigma$ [ppm/hour] &
$\mathrm{SNR}_{\mathrm{hour}}$ & $\tilde{T}_{\mathrm{max}}$\,[hours] & $\mathrm{P}(\mathrm{transit})$ \\ [0.5ex] % inserts table
%heading
\hline
Alp Cen B b	&
$1.127 \pm 0.096$ & % mass
$[0.82,0.93,1.04,1.16,1.34]$ & % radius
%$0.91_{-0.03}^{+0.03}$ & % rstar
$[0.07,0.09,0.11,0.14,0.19]$ & % depth
%1.33 &
%24.5 &
$[1.7,2.2,2.8,3.6,4.8]$ & % SNR
$2.42_{-0.08}^{+0.08}$ & % tmax
10.1\% \\
GJ 581 e	&
$1.95 \pm 0.22$ &
$[0.75,1.04,1.23,1.45,2.24]$ &
%$0.299_{-0.010}^{+0.010}$ &
$[0.5,1.0,1.4,2.1,4.7]$ &	
%10.57 &
%47.1 &
$[8,17,23,35,78]$ &
$1.19_{-0.05}^{+0.05}$ & % tmax
4.0\% \\
HD 20794 c	&
$2.36 \pm	0.43$ &
$[0.75,1.06,1.32,1.81,2.69]$ &
%$1.227_{-0.045}^{+0.045}$	&
$[0.03,0.06,0.10,0.18,0.41]$ &
%4.26 &
%24.5 &
$[0.8,1.5,2.4,4.5,10.0]$ &
$8.2_{-0.3}^{+0.4}$ & % tmax
2.8\% \\
HD 20794 b	&
$2.70 \pm 0.31$ &
$[0.76,1.09,1.42,1.99,2.88]$ &
%$1.227_{-0.045}^{+0.045}$ &
$[0.03,0.07,0.11,0.22,0.47]$ &
%4.26 &
%24.5 &
$[0.8,1.6,2.8,5.4,11.5]$ &
$6.3_{-0.3}^{+0.3}$ & % tmax
4.7\% \\
HD 85512 b	&
$3.62 \pm 0.44$ &
$[0.85,1.22,1.70,2.42,3.46]$ &
%$0.882_{-0.081}^{+0.081}$ &
$[0.08,0.16,0.32,0.65,1.37]$ &
%7.67 &
%24.5 &
$[2.1,4.5,8.9,18.4,38.5]$ &
$6.4_{-0.7}^{+0.7}$ & % tmax
1.6\% \\
$\vdots$ & $\vdots$ & $\vdots$ & $\vdots$ & $\vdots$ & $\vdots$ & $\vdots$ \\ [1ex]
\hline %inserts single line
\label{tab:final}
\end{tabular}
\end{table*}

For each planet, we also recorded the most probable classification outputted
from \forecaster\ (Terran, Neptunian, Jovian or Stellar). Amongst the 468
RV planets studied, we find zero instances of a Stellar classification, which
is to be expected, 354 Jovians, 112 Neptunians and 2 Terrans (Alpha Centauri A
b and GJ 581 e). Amongst the Neptunian worlds, we split them into groups based
by dividing about 10\,$M_{\oplus}$ in terms of maximum a-posteriori reported
mass on EOD, defining a subset of mini-Neptunes (28 planets) and Neptunes (84
planets).

\begin{figure}
\begin{center}
\includegraphics[width=8.4cm,angle=0]{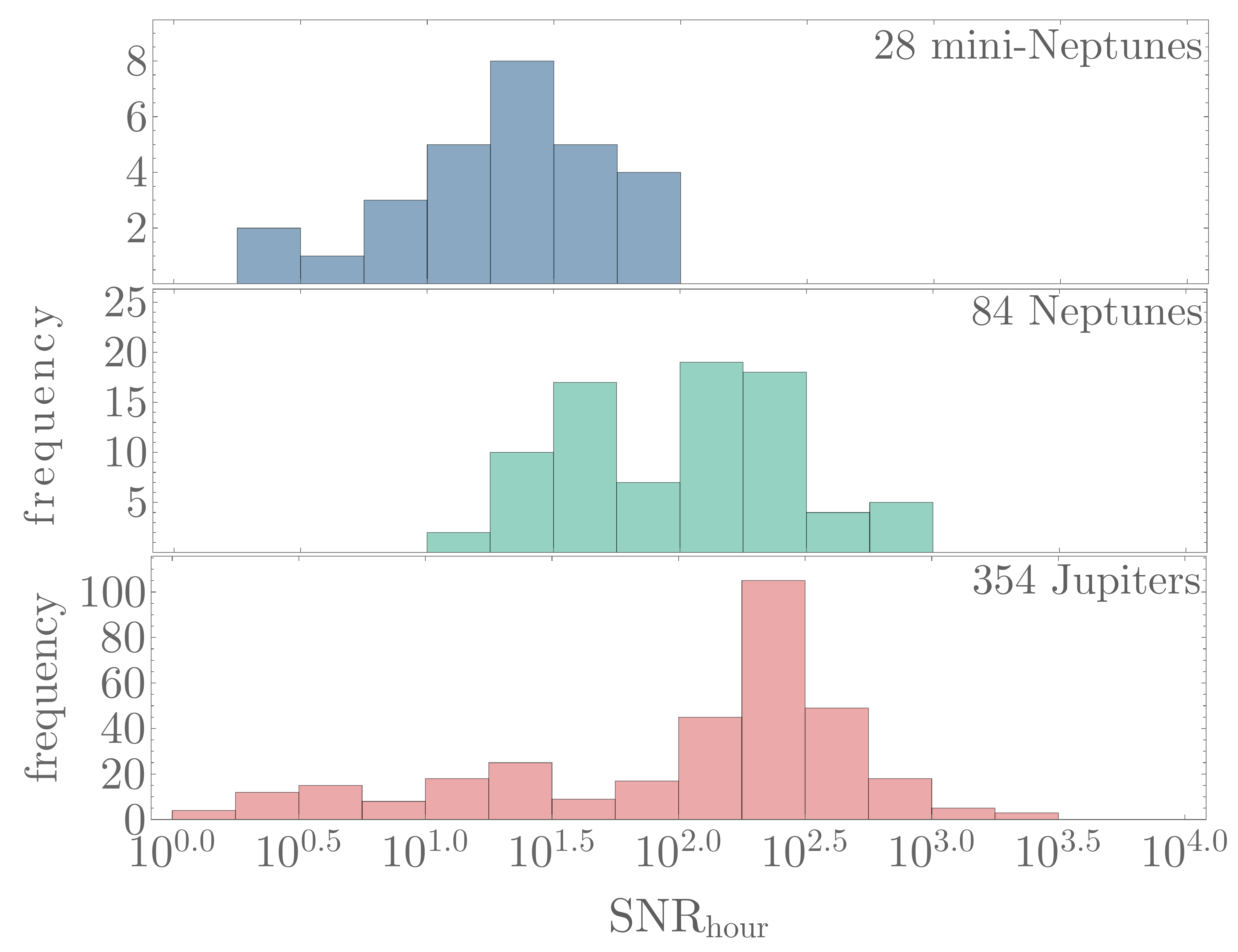}
\caption{
Histograms of the $\mathrm{SNR}_{\mathrm{hour}}$ (in log space) for the three
broad class of planets considered in this work. The majority of the RV planets
would be detectable in transit using CHEOPS with a single transit observation,
assuming the stars are not highly active.
}
\label{fig:histo}
\end{center}
\end{figure}

In Figure~\ref{fig:histo}, we show three histograms for each type of planet's
maximum a-posteriori forecasted $\mathrm{SNR}_{\mathrm{hour}}$. Given
that even short-period planetary transits typically last for a couple of hours,
the high $\mathrm{SNR}_{\mathrm{hour}}$ values seen in Figure~\ref{fig:histo}
imply that the vast majority of these RV planets are expected to be detectable
with CHEOPS with a single event. Specifically, we find a maximum a-posteriori
$\mathrm{SNR}_{\mathrm{hour}}$ exceeding 10 for 22 of the 28 mini-Neptunes, 82
of the 84 Neptunes and 294 of 354 Jovians. However, we highlight that our
$\mathrm{SNR}_{\mathrm{hour}}$ calculations assume no intrinsic photometric
noise for the parent stars, which may in some case significantly affect the
estimates given (discussed further in Section~\ref{sec:discussion}).

For the two Terrans, GJ 581 e is expected to be easily detectable (assuming it
transits) at $\mathrm{SNR}_{\mathrm{hour}}\sim$20 but Alpha Cen B b would be
more challenging at $\mathrm{SNR}_{\mathrm{hour}}$ of a few, requiring likely
more than a single transit observation. However, we point out that transits of
both of these high-profile planets has already been investigated, with transits
generally excluded for both \citep{dragomir:2012,demory:2015}.

\subsection{CHEOPS vs \kepler\ planets}

\begin{figure}
\begin{center}
\includegraphics[width=8.4cm,angle=0]{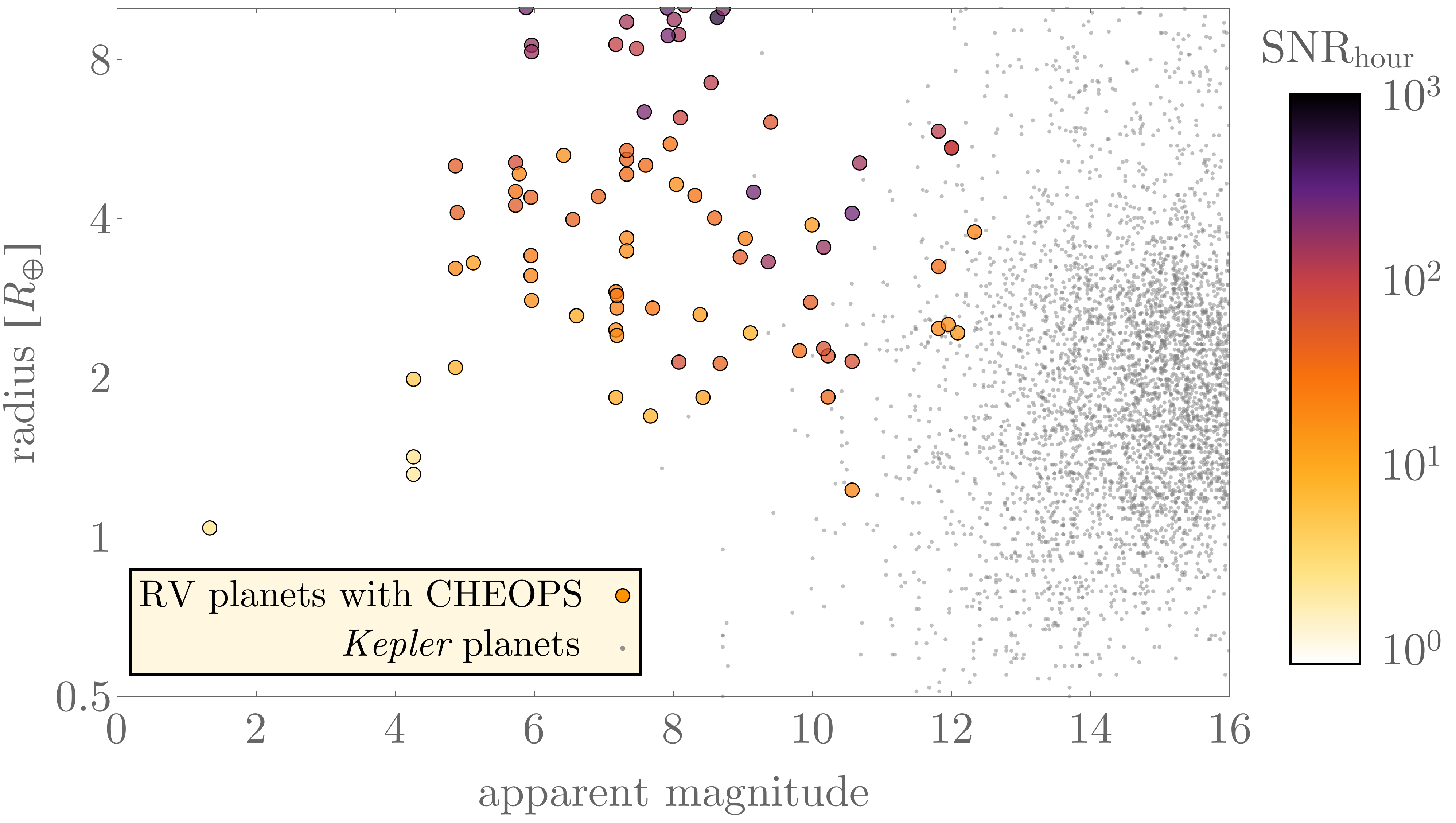}
\caption{
Radii of the \kepler\ planets (gray) as function of their apparent magnitude.
For comparison, we show the RV planets where the $\mathrm{SNR}_{\mathrm{hour}}$
is provided using the colour scale, assuming the CHEOPS-like SNRs derived in
this work. As expected, CHEOPS indeed has the capability to provide transiting
planets orbiting far brighter stars than that of \kepler.
}
\label{fig:kepler}
\end{center}
\end{figure}

CHEOPS is not a wide-field transit survey and thus should not be expected to find
remotely near as many planets as that discovered by \kepler. However,
transiting planets discovered by CHEOPS have the potential to be significantly
brighter than those of \kepler, an argument which originally motivated the CHEOPS
project \citep{broeg:2013}. To investigate this, Figure~\ref{fig:kepler} shows
the apparent magnitude (treating $V$-band and the \kepler\ bandpass as being
approximately equal) of the \kepler\ planets as a function of their size,
versus the RV planets, where $\mathrm{SNR}_{\mathrm{hour}}$ is computed as
described in Section~\ref{sub:SNR}.

Figure~\ref{fig:kepler} confirms that if CHEOPS targets the bright RV planets,
it should be expected to deliver a subset of significantly brighter transiting
planets than those found by \kepler. However, we also stress that the planet
yield will almost certainly be far smaller, and the smallest possible detectable
planet will be larger than the smaller \kepler\ worlds, owing to the fact RV
detections do not currently probe into the sub-Earth mass regime reliably.

\subsection{Validating the predictions}

The \forecaster\ model from \citet{chen:2017a} has been subject to extensive
testing, particularly in the original paper but also that of
\citet{chen:2017b}. Nevertheless, to validate our predictions we subject to
it another test by extracting the RV planets in our sample for which the
planetary radius has been directly measured via a transit detection.

\begin{figure}
\begin{center}
\includegraphics[width=8.4cm,angle=0]{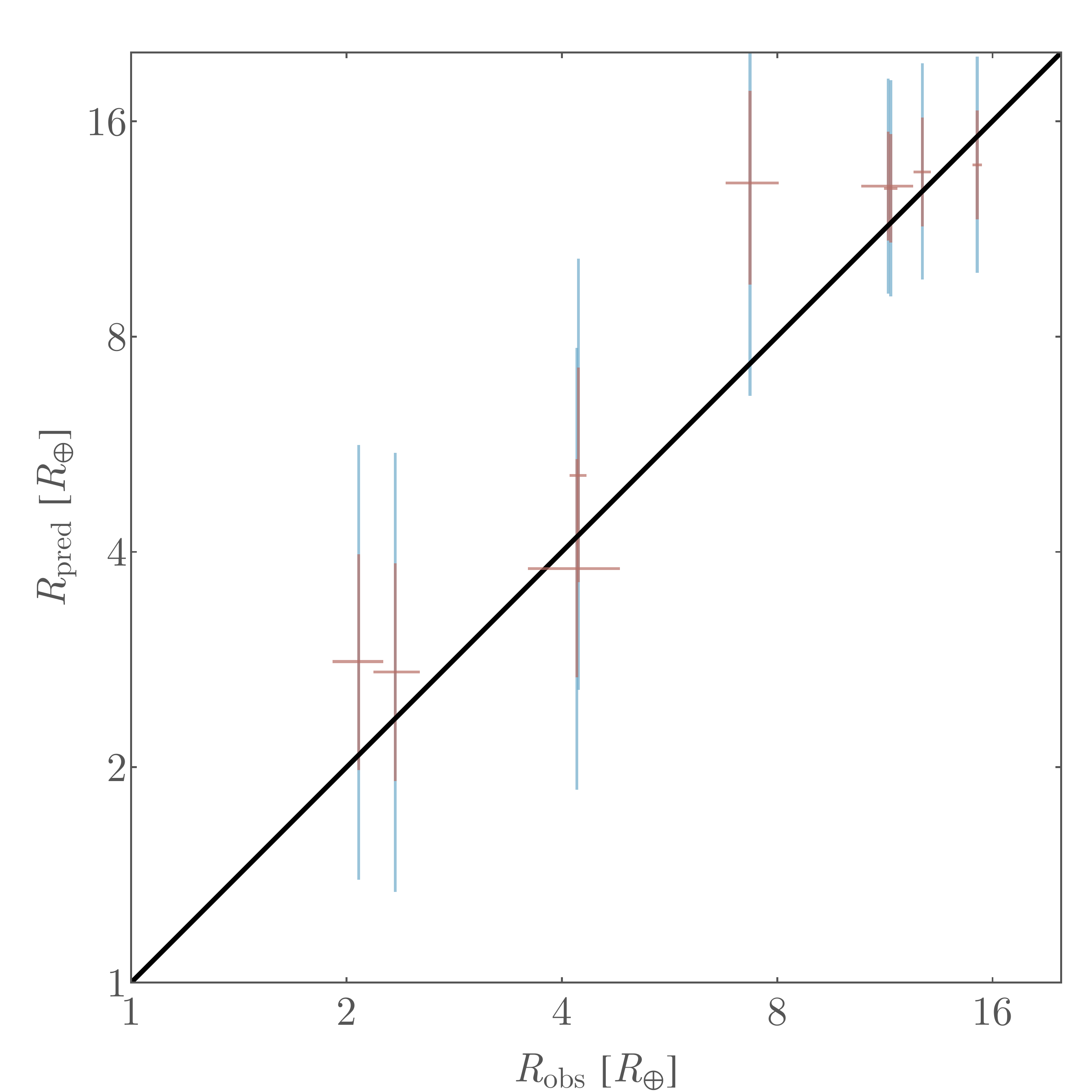}
\caption{
A comparison of predicted planetary radii from this work using the
\forecaster\ package of \citet{chen:2017a} and reported measured planetary
radii. Only nine planets in our sample have a measured radius from transits.
Blue denotes the 2\,$\sigma$ confidence region and red the 1\,$\sigma$.
}
\label{fig:comp}
\end{center}
\end{figure}

There are only nine such cases in our sample (HD 189733b, HD 209458b, HD 80606b,
GJ 436b, GJ 3470b, HD 17156b, HD 149026b, HD 97658b, 55 Cnc e). We compare the
predictions versus the observations in Figure~\ref{fig:comp}, which reveals
that our predictions are fully compatible with the observed values. Only
one of our predicted radii falls outside of the 1\,$\sigma$ credible interval
(HD 149026b), yet this object falls within the 2\,$\sigma$ prediction.
This is not statistically surprising, since amongst a sample of nine objects it
should not be generally expected that all nine fall within a 68.3\% probability
interval.

\section{Discussion}
\label{sec:discussion}

In this work, we have predicted the planetary radii, transit depths and
associated signal-to-noises (SNRs) using CHEOPS \citep{broeg:2013} for
468 planets listed on the ``Exoplanet Orbit Database''
(\href{http://exoplanets.org}{EOD}; \citealt{han:2014}) as having been
discovered using the radial velocity (RV) method. Our SNR values assume
one-hour of cumulative integration time, no intrinsic stellar noise, employ a
CHEOPS noise model dependent on spectral type and apparent magnitude, and
accounts for instrument saturation. This fiducial choice of a one-hour
integration time is a general value which may be easily transformed to account
for the number of transits observed, the observational duty cycle of each
transit (e.g. due to CHEOPS' orbit) and the (unknown) transit duration
(see Equation~\ref{eqn:SNRreal}). Predictions for planetary radii are based
using the empirical, probabilistic framework known as \forecaster\ presented by
\citet{chen:2017a}.

We have verified that our predictions are consistent with the observed
planetary radii in nine cases where the RV planets have been established to
transit. Amongst the other planets, we predict that CHEOPS will be able to
detect transits at $\mathrm{SNR}_{\mathrm{hour}}>10$ for the vast majority,
assuming that they have the correct geometry and our assumptions hold true.
The greatest unknown quantity in our model is the potential for intrinsic
photometric variability of the parent stars, which in optical bandpasses has
been demonstrably shown to a major impediment to transit discovery if present
\citep{kipping:2017}. For this reason, it may be worthwhile for CHEOPS
to conduct preliminary photometric observations of planned targets during
quiet times, in order to ascertain the possibility of activity.

Using the classifications made by \forecaster, we predict that only two
of the RV planets considered are most likely Terran (Alpha Cen A b and
GJ 581 e), yet both have already been established to be unlikely to
transit \citep{dragomir:2012,demory:2015}. Amongst the 92 mini-Neptunes
and 20 Neptune-sized planets, we find typical transit probabilities
of order 5\% and thus predict numerous discoveries from CHEOPS with a
suitable observing strategy.

The main objective of this work is to aid the CHEOPS team in predicting
which RV planets are most suitable for follow-up efforts in a quantifiable
manner. Aside from pursuing RV planets, we highlight that it may also
be fruitful to target known planetary systems where additional planets
are confidently predicted using machine learning methods, such as that
presented in \citet{lam:2017}. Undoubtedly targets-of-opportunity will
arise before the expected launch in 2018, yet a library of pre-computed
forecasted SNRs should aid the team in scheduling observations amongst
these other opportunities. Whilst this work has certainly had CHEOPS
directly in mind due to the unique strategy planned, the predictions
for the transit depths are not specific to CHEOPS and thus may also
benefit other transit surveys.

Our predictions establish that CHEOPS should find transiting planets
orbiting far brighter target stars than that of \kepler. For example,
amongst the 104 Neptunian RV planets with satisfying 
$\mathrm{SNR}_{\mathrm{hour}}>10$, 86 of them will be orbiting stars brighter
than $V=10$, providing exciting opportunities for atmospheric characterization
both from the ground and space. To aid the community in planning both
the CHEOPS observing strategy and potential follow-up, we make our
predictions publicly available at \httplink.

\section*{Acknowledgments}

Special thanks to David Ehrenreich and the CHEOPS team for their
assistance with the noise model. We thank the anonymous reviewer
for their helpful feedback in clarifying the assumptions of our
work.
DMK acknowledges support from NASA grant NNX15AF09G (NASA ADAP Program).
This research has made use of the \forecaster\ predictive
package by \citet{chen:2017a}, the {\tt corner.py} code by Dan
Foreman-Mackey at
\href{http://github.com/dfm/corner.py}{github.com/dfm/corner.py} and
the Exoplanet Orbit Database and the Exoplanet Data Explorer at
exoplanets.org.

%\appendix
%
%\input{lambda.tex}

\bsp
\label{lastpage}
\end{document}